\input{aipcheck}

\documentclass[
  ,draft            
  ,numberedheadings 
  ]
  {aipproc}

\layoutstyle{6x9}

\begin{document}

\begin{flushright} 
{FTUV-2012-08-28 \hskip 1cm UAB-FT-719
\hskip 1 cm IFIC-12-61   } 
\end{flushright}

\vspace{-0.5cm}\title{MC generator TAUOLA: implementation of Resonance Chiral Theory for two and three meson modes. Comparison with  experiment}


\author{O. Shekhovtsova}{
  address={IFIC, Universitat de Val\`encia-CSIC,  Apt. Correus 22085,   E-46071, Val\`encia, Spain\footnote{Speaker}
}
}

\author{I. M. Nugent}{
  address={RWTH Aachen University, III. Physikalisches Institut B, Aachen, Germany}
}

\author{T. Przedzinski}{
  address={The Faculty of Physics, Astronomy and Applied Computer
Science, \\ Jagellonian University, Reymonta 4, 30-059 Cracow, Poland}
  ,altaddress={CERN PH-TH, CH-1211 Geneva 23, Switzerland} 
}

\author{P. Roig}{
  address={Grup de F\'{\i}sica Te\`orica, Institut de F\'{\i}sica d'Altes Energies, 
Universitat Aut\`onoma de Barcelona, E-08193 Bellaterra, Barcelona, Spain} 
}

\author{Z. Was}{
  address={ Institute of Nuclear Physics, PAN,
        Krak\'ow, ul. Radzikowskiego 152, Poland}
  ,altaddress={CERN PH-TH, CH-1211 Geneva 23, Switzerland} 
}

\begin{abstract}
We present a partial upgrade of the Monte Carlo event generator 
TAUOLA with the two and three hadron decay modes using the theoretical 
models based on Resonance Chiral Theory.
These modes
account for 88\% of total hadronic width of the tau meson.  First results 
of the model parameters have been
obtained using BaBar data for 3$\pi$ mode.

\end{abstract}

\maketitle


\section{Introduction}
\vspace{-0.3cm}
Tau lepton is the only lepton that decays into hadrons.  From the perspective of low   energies, 
 as explored at B-factories, BaBar \cite{babar}  and Belle \cite{belle} experiments, 
hadronic decay modes of tau lepton provide an excellent laboratory for modeling hadronic interactions at the energy scale of about 1-2 GeV, 
where neither perturbative QCD methods nor chiral lagrangians are expected to work with a good precision \cite{Braaten:1990ef}.
 From the perspective of high energy experiments such as those at LHC, good understanding of tau leptons properties contributes 
important ingredients of new physics signatures. 
With the discovery of a new particle around the mass of 
125-126GeV \cite{higgs:2012}, tau decays are an important 
ingredient for determining if this is the Standard Model Higgs.
This is especially pertinent since CMS has reported a deficit in the 
number of fermion decays from the new particle relative to the Standard 
Model Higgs Prediction.

The first version of the program TAUOLA \cite{Jadach:1993hs} was written in the 90's. 
TAUOLA is Monte Carlo event generator which simulates tau decays for 
both the leptonic and hadronic decays modes. The hadronic currents 
implemented in TAUOLA are based on resonance dominance model (RDM) \cite{Kuhn} 
\footnote{Based on \cite{Jadach:1993hs}  Cleo and Aleph collaborations  developed their own versions, non published, 
which differ from one  another by the intermediate resonance states
 because of different detector sensitivity.}.
In the framework of RDM the hadronic current for a three-pseudoscalar 
decay mode is a sum of weighted products of the Breit-Wigner functions.  It is demonstrated in \cite{GomezDumm:2003ku} 
that this approach is able to reproduce only LO $\chi$PT result. 
Moreover, the model was not sufficient to describe the 
Cleo $KK\pi$ data \cite{Coan:2004ep}. This resulted in the Cleo collaboration reshaping the 
model by introducing two ad-hoc parameters that spoilt the QCD 
normalization for Weiss-Zumino contribution.
The parameters are obtained fitting to data.
 However, before making conclusion that the Wess-Zumino anomaly normalization is spoilt it should be checked 
whether on oversimplified theoretical approximation, like RDM, was applied.

The alternative approach based on the Resonance Chiral Theory (R$\chi$T) \cite{rcht} was proposed in  \cite{GomezDumm:2003ku}.  
The computations done within R$\chi$T are able to
reproduce the low-energy limit of $\chi$PT at least up to NLO and demonstrate the right falloff 
in the high energy region \cite{Roig:talk}.
The hadronic currents for the two and three meson modes ($\pi\pi$, $K\pi$, $K K$, $\pi\pi\pi$ and $K K\pi$) 
calculated in the framework of R$\chi$T 
were implemented in TAUOLA \cite{Shekhovtsova:2012ra}.  
The tar-ball version of the code is available at web-page \cite{web:RChL}. 

\section{Results for two and three meson final states}
\vspace{-0.3cm}
It is of utmost importance to implement the theory in a form that is as useful for general applications as possible. 
Therefore the corresponding hadronic currents ($J^\mu$) in the upgraded version of TAUOLA have been   written 
in the most general form compatible with Lorentz invariance. 
For $\tau$ decay channels with two mesons ($h_1(p_1)$ and $h_2(p_2)$)
\begin{equation}\label{eq:twomes}
J^\mu  = N \bigl[ (p_1 - p_2 - \frac{\Delta_{12}}{s}(p_1 +p_2))^\mu F^{V}(s) + \frac{\Delta_{12}}{s}((p_1 + p_2)^\mu F^{S}(s) \bigr],
\end{equation}
where $\Delta_{12} = m_1^2 -m_2^2$, $s = (p_1 +p_2)^2$.
For the final state of three pseudoscalars, with momenta $p_1$, $p_2$ and $p_3$, Lorentz invariance determines the decomposition 
\begin{eqnarray}
J^\mu &=N &\bigl\{T^\mu_\nu \bigl[ c_1 (p_2-p_3)^\nu F_1  + c_2 (p_3-p_1)^\nu
 F_2  + c_3  (p_1-p_2)^\nu F_3 \bigr]\nonumber\\
& & + c_4  q^\mu F_4  -{ i \over 4 \pi^2 F^2}      c_5
\epsilon^\mu_{.\ \nu\rho\sigma} p_1^\nu p_2^\rho p_3^\sigma F_5      \bigr\},
\label{eq:fiveF}
\end{eqnarray}
where:  $T_{\mu\nu} = g_{\mu\nu} - q_\mu q_\nu/q^2$ denotes the transverse
projector, $q^\mu=(p_1+p_2+p_3)^\mu$ is the momentum of the hadronic system, 
$F$ is the pion decay constant in the chiral limit.

The decomposition (\ref{eq:twomes}) and (\ref{eq:fiveF}) is the most general one, model-dependence is included in 
the hadronic form factors ($F_V$, $F_S$ as well as $F_i$, i = 1...5).
The hadronic form factors  calculated within R$\chi$T can be written as
\begin{equation}
F_I = F_I^\chi + F_I^R + F_I^{RR}
\end{equation}
where $F_I^{\chi}$ is the chiral contribution, $F_I^R$ is the one resonance contribution and $F_I^{RR}$ is the double-resonance part. 
The explict form of the functions $F_i$ for 3$\pi$ and $KK\pi$ modes can be found in  
\cite{Shekhovtsova:2012ra}, Section 2, as well as in  \cite{GomezDumm:2003ku,Dumm}. 
The theoretical results  for the hadronic currents  were obtained 
in the isospin limit ($m_{\pi}=(2m_{\pi^+}+m_{\pi^0})/3$, $m_K=(m_{K^+}+m_{K^0})/2$), except for the two pion and two kaons modes. 
In the phase space generation, the differences between neutral
and charged pion and kaon masses is taken into account, i.e. physical values are chosen. This has to be done to obtain proper kinematic configurations. 
The  model parameters, more specifically the masses of the resonances and the coupling constants, were fitted to Aleph data, requiring 
correct high-energy behaviour of the related form factors, see Appendix C in \cite{Shekhovtsova:2012ra}.

To check stability of multidimensional integration in TAUOLA the MC results were compared with the semi-analytical ones 
(Gauss integration of analytical results was used). 
The difference  between MC prediction  and semi-analytical results  for the partial decay width is less than $0.02\%$ for  all channels.
Both differential spectra  and numerical tests are collected at 
\cite{web:RChL}.

In Table \ref{tab:bench}, the partial decay width values from PDG \cite{Nakamura:2010zzi} 
are compared with our results obtained with isospin-averaged pseudoscalar masses and with the physical ones. 
Comparison of the last two columns illustrates numerical effect of physical masses.
The difference between the MC result and PDG one is $2\% - 24\%$ depending on the channel.
As expected, that agreement is not good because only minimal attempts 
on adjusting to
the model parameters have been applied for the comparison with BaBar and 
Belle data.  The next section
presents the first step toward this direction.
\begin{table}
\vspace{-0.5cm} 
{ \begin{tabular}{|r| c| c| c|} 
\hline
Channel & \multicolumn{3}{c|}{Width, [GeV]} \\
\cline{2-4}
    & PDG  & Equal masses &   Phase space\\  
   &  &  &  with masses \\  
\hline
{$ \pi^-\pi^0 \; \;\; \;$}  &      {($5.778 \pm 0.35\%)\cdot 10^{-13}$} &  {($5.2283 \pm 0.005\%)\cdot10^{-13}$} &  {$(5.2441\pm 0.005\%)\cdot 10^{-13}$} \\ 
{$ K^-\pi^0 \; \;\; \;$}   &       {($9.72\;\pm 3.5\%\;)\cdot 10^{-15}$} &  {($8.3981 \pm 0.005\%)\cdot10^{-15}$} &  {$(8.5810\pm 0.005\%)\cdot 10^{-15}$} \\ 
{$ \pi^-\bar K^0 \; \;\; \;$}   &  {($1.9\;\;\; \pm 5\%\;\;\;)\cdot 10^{-14}$} &  {($1.6798 \pm 0.006\%)\cdot10^{-14}$} &  {$(1.6512\pm 0.006\%)\cdot 10^{-14}$} \\ 
{$ K^-K^0 \; \;\; \;$}     &       {($3.60\; \pm 10\%\;\;)\cdot 10^{-15}$} &  {($2.6502 \pm 0.007\%)\cdot10^{-15}$} &  {$(2.6502 \pm 0.008\%)\cdot 10^{-15}$} \\ 
 {$  \pi^-\pi^-\pi^+$} &           {($2.11\; \pm 0.8\%\;\;)\cdot 10^{-13}$} &  {($ 2.1013\pm 0.016\%)\cdot10^{-13}$} &   {$(2.0800\pm 0.017\%)\cdot 10^{-13}$}  \\ 
{$  \pi^0\pi^0\pi^-$}  &           {($2.10\; \pm 1.2\%\;\;)\cdot 10^{-13}$} &  {($ 2.1013\pm 0.016\%)\cdot10^{-13}$} &  {$(2.1256\pm 0.017\%)\cdot 10^{-13}$}\\ 
 {$  K^-\pi^-K^+$} &   {($3.17\; \pm 4\%\;\;\;)\cdot 10^{-15}$} &  {($3.7379 \pm 0.024\%)\cdot10^{-15}$} &   {$(3.8460\pm 0.024\%)\cdot  10^{-15}$}  \\ 
{$  K^0\pi^-\bar{K^0}$}  &         {($3.9\;\; \pm 24\%\;\;)\cdot 10^{-15}$} &  {($3.7385 \pm 0.024\%)\cdot10^{-15}$} &  {$(3.5917\pm 0.024\%)\cdot 10^{-15}$}\\
{$  K^-\pi^0 K^0$} &               {($3.60\; \pm 12.6\%\;\;)\cdot 10^{-15}$} &  {($2.7367\pm 0.025 \%)\cdot10^{-15}$} &  {$(2.7711 \pm 0.024\%)\cdot 10^{-15}$}\\
\hline
\end{tabular} 
}  
\vspace{-0.5cm}
\caption{The $\tau$ decay partial widths. 
The PDG value \cite{Nakamura:2010zzi} (2nd collumn) is compared with numerical results 
from  \texttt{TAUOLA} with the R$\chi$T currents: in the isospin limit for pseudoscalar masses (3rd collumn), using  physical masses (4th collumn).
%
} \label{tab:bench}
\end{table}

\section{Fit for three pion mode to BaBar data}
\vspace{-0.3cm}
The main problem with upgrading the MC simulation 
of $\tau$ decays is a lack of the published spectrum. Currently, only 
the two pion modes \cite{belle} and three pseudoscalar modes
\cite{Nugent:2009zz} are published.

\begin{figure}\label{fig:Ian}
\vspace{-0.5cm}
\begin{tabular}{c}
\includegraphics[height=.27\textheight]{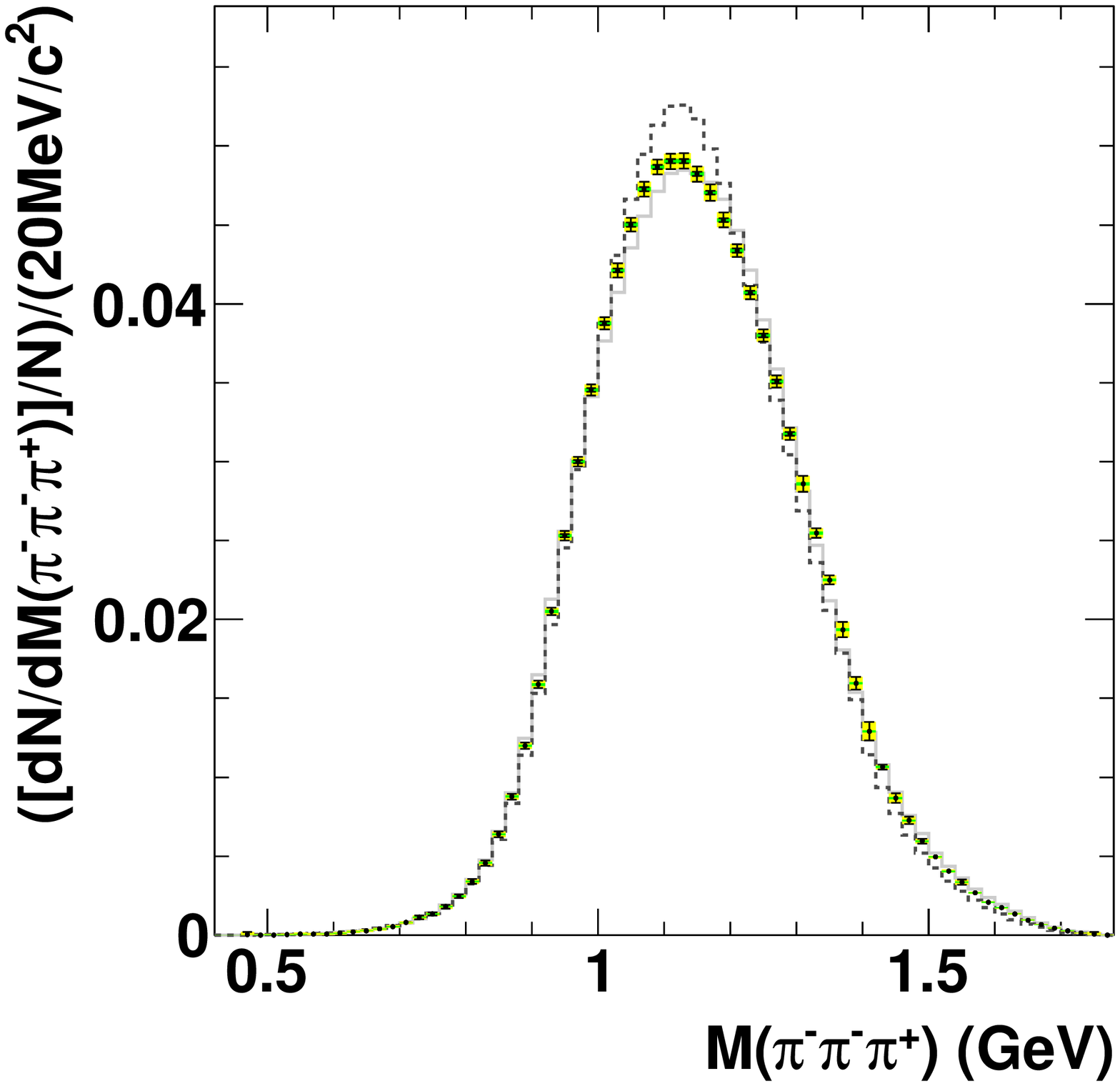}
\includegraphics[height=.27\textheight]{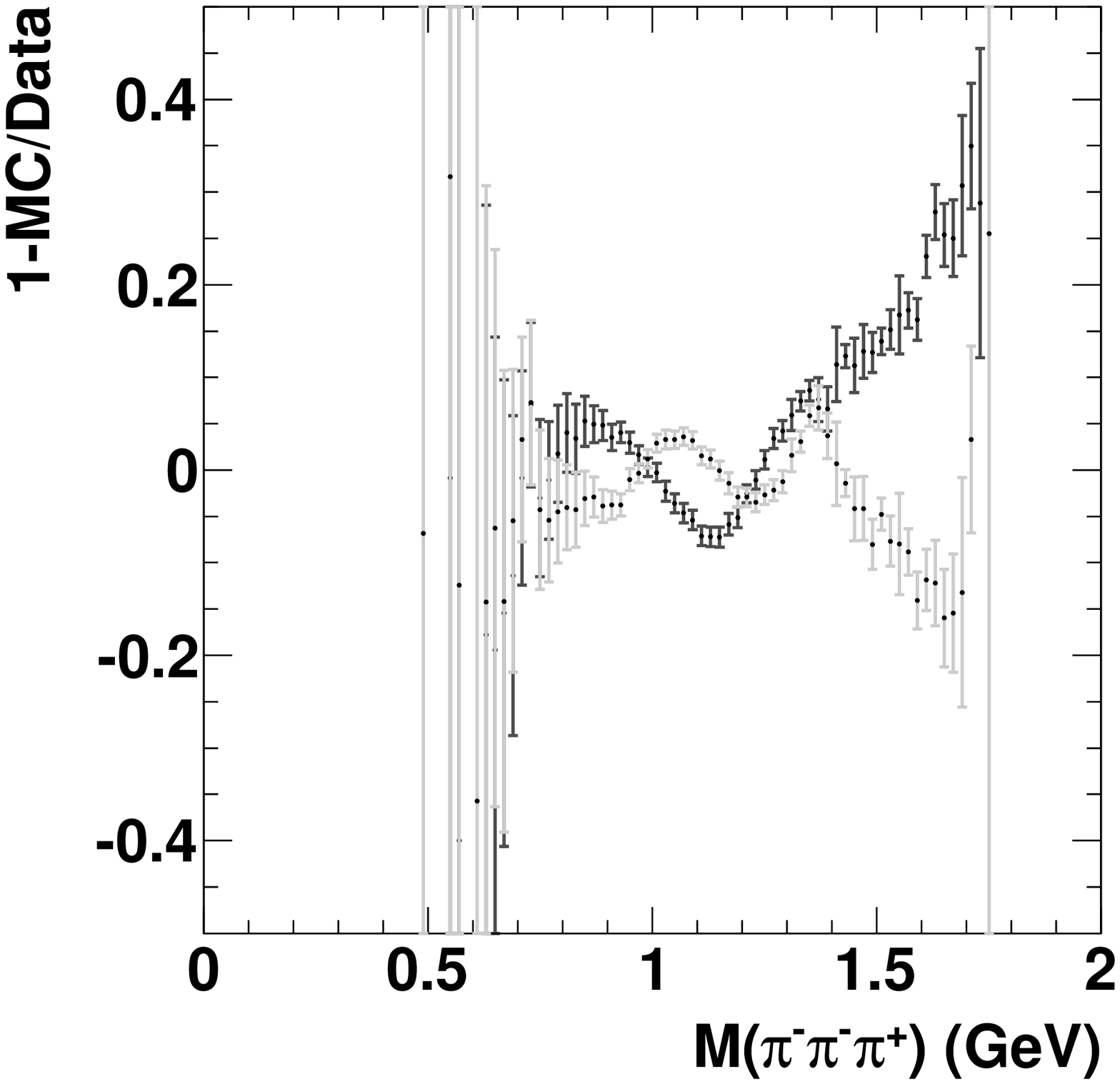}
\vspace{-0.2cm}
\\ \vspace{-0.2cm}
\includegraphics[height=.27\textheight]{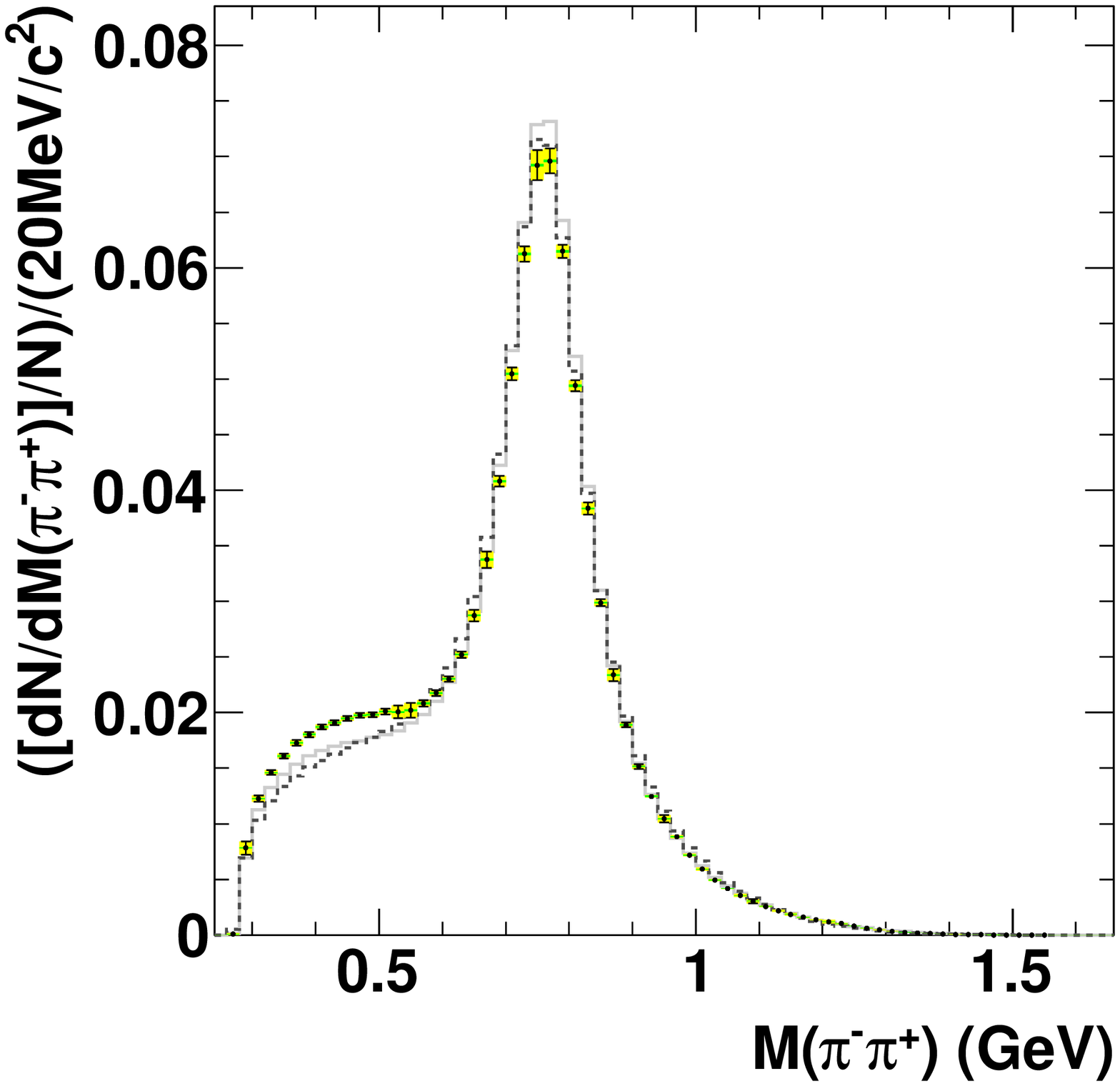}
\includegraphics[height=.27\textheight]{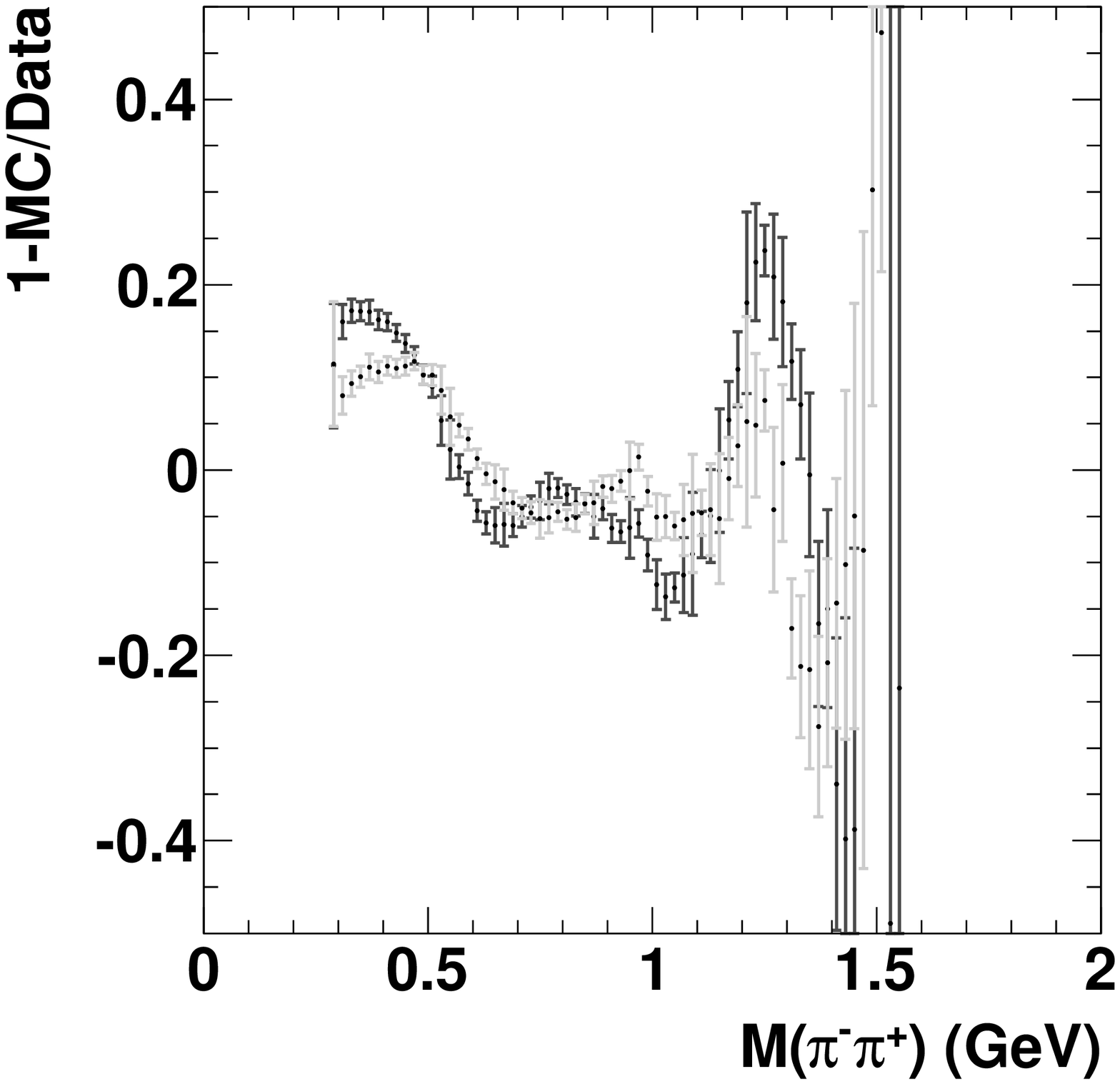}
\end{tabular}
\caption{Invariant $\pi^+\pi^-\pi^-$ (up) and  
$\pi^+\pi^-$ (down) mass distributions: the plots on the left-hand side correspond to the differential decay distribution, 
the ones on the right-hand side to plot ratios between MC and BaBar data.
Lighter grey histograms are for R$\chi$T parametrization, darker grey is for {\texttt CLEO} one.}
\end{figure}

The results from a preliminary fit to the three pion mode ($\pi^+\pi^-\pi^-$) \cite{Nugent:2009zz}  can be 
seen in Fig.  \ref{fig:Ian} and Table \ref{tab:fit}.
Both the three particle and the $\pi^+\pi^-$ invariant 
mass distributions have been considered. Disagreement about 12\% level is visible
in the low energy region of two particle invariant mass distribution whereas for the three invariant mass spectrum the difference between MC and data is less than 7\%.
However, the R$\chi$T parametrization is in better agreement with BaBar data than {\tt CLEO} one, see Fig. \ref{fig:Ian}.

\begin{table}
\vspace{-0.3cm}
\begin{tabular}{|l|l|l|l|l|l|l|l|}
\hline
  & $M_{\rho'}$& $\Gamma_{\rho'}$& $M_{a_1}$& $F$& $F_V$& $F_{A}$& $\beta_{\rho'}$\\
\hline
Min. & 1.44 & 0.32 & 1.00    & 0.0920 & 0.12 & 0.1 & -0.36\\
\hline
Max & 1.48 & 0.39 & 1.24    & 0.0924 & 0.24 & 0.2 & -0.18 \\
\hline
Default & 1.453 & 0.4 & 1.12    & 0.0924 & 0.18 & 0.149 & -0.25 \\
\hline
Fit \tablenote{to\cite{Nugent:2009zz}, $\chi^2/ndf = 2262.12/132$} & 1.4302 & 0.376061 & 1.21706    & 0.092318 & 0.121938 & 0.11291 & -0.208811\\
\hline
\end{tabular}
\vspace{-0.5cm}
\caption{Numerical values of the R$\chi$T parameters fitted to BaBar data for three pion mode}
\label{tab:fit}
\end{table}

The fit was done taking into account only $P$-wave contribution of two pion system. 
As suggesting in \cite{Shibata:2002uv} the discrepance in the low mass region could be described using a contribution 
from the scalar particle, $S$-wave contribution.
We expect that inclusion of $S$-wave contribution \cite{Roig:talk} 
 will improve the value of $\chi^2$. 
The values in the fifth row of Table \ref{tab:fit}
are only the preliminary results. They will not necessarily correspond to the minimum of $\chi^2/ndf$
of the final fit. Work is in progress.

%

\section{Conclusion}
\vspace{-0.3cm}
The theoretical results for the hadronic currents of two and three pseudoscalar modes, namely, $\pi \pi$, $K\pi$, $K K$, $\pi\pi\pi$ and $K K \pi$, 
 in the framework of R$\chi$T
have been implemented in TAUOLA. These modes, together with the one-meson decay modes, represent more than 88\% of the hadronic width of tau.   
R$\chi$T is a more controlled QCD-based model than the usual used Breit-Wigner parametrizations. 
However, before making conclusion about validity of the model the theoretical results have to be confronted with the experimental data, 
which requires fit of the model parameters. 
Now that the technical work on current installation is complete,
 the work on fits is in progress in collaboration with 
 theoreticians and experimentalists.

At LHC tau decays are only used for identification and are not used 
to study their dynamic. However, the dynamics of tau decays are 
important for both modeling the decays and -therefore the reconstruction 
and identification-  and for measuring the polarization of tau decays. 
Therefore, an upgrade to the TAUOLA based on the BaBar and Belle results 
on tau decays is urgently needed for LHC.


\begin{theacknowledgments}
 \vspace{-0.3cm}
This research  was supported by a Marie Curie Intra European Fellowship within 
the 7th European Community Framework Programme 
(O.S.) and by Alexander von Humboldt Foundation (I.N.),
 by  the Spanish Consolider Ingenio 2010 Programme 
CPAN (CSD2007-00042) and by  MEC (Spain) under Grants FPA2007-60323, FPA2011-23778 (O.S. and P.R.) and FPA2011-25948 (P.R.) and in part by the funds of Polish National Science Centre under decision DEC-2011/03/B/ST2/00107 (T.P., Z.W.). 
\end{theacknowledgments}

\bibliographystyle{aipproc}

\end{document}